\documentclass[aip,amsmath,amssymb,reprint,numerical]{revtex4-1}

\usepackage[utf8]{inputenc}
\usepackage[T1]{fontenc}
\usepackage{mathptmx}         % Times font
\usepackage{etoolbox}
\usepackage{physics}
\usepackage{braket}
\usepackage{tikz}
\usetikzlibrary{arrows}

\usepackage{silence}
\usepackage{hyperref}         % always last

\makeatletter
\def\@email#1#2{%
 \endgroup
 \patchcmd{\titleblock@produce}
  {\frontmatter@RRAPformat}
  {\frontmatter@RRAPformat{\produce@RRAP{*#1\href{mailto:#2}{#2}}}\frontmatter@RRAPformat}
  {}{}
}%
\makeatother
\begin{document}

\preprint{AIP/123-QED}

\title{Gauge-Invariant Long-Wavelength TDDFT Without Empty States: From Polarizability to Kubo Conductivity Across Heterogeneous Materials}

\author{Christian Tantardini*}
\email{christiantantardini@ymail.com}
\affiliation{Center for Integrative Petroleum Research, King Fahd University of Petroleum and Minerals, Dhahran 31261, Saudi Arabia}

\author{Quentin Pitteloud}
\affiliation{Hylleraas center, Department of Chemistry, UiT The Arctic University of Norway, PO Box 6050 Langnes, N-9037 Troms\o, Norway.}

\author{Boris Yakobson}
\affiliation{Department of Materials Science and NanoEngineering, Rice University, Houston, Texas 77005, United States of America.}

\author{Martin Peter Andersson}
\affiliation{Center for Integrative Petroleum Research, King Fahd University of Petroleum and Minerals, Dhahran 31261, Saudi Arabia}

\date{\today}

\begin{abstract}
Electromagnetic response is commonly computed in two languages: length-gauge molecular polarizabilities and velocity-gauge (Kubo) conductivities for periodic solids. We introduce a compact, gauge-invariant bridge that carries the same microscopic inputs—transition dipoles and interaction kernels—from molecules to crystals and heterogeneous media, with explicit SI prefactors and fine-structure scaling via \(\alpha_{\rm fs}\). The long-wavelength limit is handled through a reduced dielectric matrix that retains local-field mixing, interfaces and 2D layers are treated with sheet boundary conditions (rather than naïve ultrathin films), and length–velocity equivalence is enforced in practice by including the equal-time (diamagnetic/contact) term alongside the paramagnetic current. Finite temperature is addressed on the Matsubara axis with numerically stable real-axis evaluation (complex polarization propagator), preserving unit consistency end-to-end.

The framework enables predictive, unit-faithful observables from radio frequency to ultraviolet—RF/microwave heating and penetration depth, dielectric-logging contrast, interfacial optics of thin films and 2D sheets, and adsorption metrics via imaginary-axis polarizabilities. Numerical checks (gauge overlay and optical \(f\)-sum saturation) validate the implementation. Immediate priorities include compact, temperature- and salinity-aware kernels with quantified uncertainties and \emph{operando} interfacial diagnostics for integration into multiphysics digital twins.
\end{abstract}

\maketitle

\section{Introduction}
\label{sec:intro}

Light–matter response is often described in two languages: the length-gauge polarizability used for finite molecules and the velocity-gauge (Kubo) conductivity used for periodic solids. In the ideal, complete-basis limit they are equivalent; in practice, separate implementations, incomplete bases, and the long-wavelength limit can introduce gauge inconsistencies and numerical instability \cite{RungeGross1984,Casida1995,Ullrich2012,DreuwHeadGordon2005,Kubo1956,Kubo1957,Adler1962,Wiser1963,OnidaReiningRubio2002,DresselGruner2002,AversaSipe1995,SouzaIniguezVanderbilt2002,Schuler2021,Taghizadeh2017,Taghizadeh2018,Ventura2017}. Local-field effects (microscopic charge inhomogeneities feeding back onto macroscopic optical constants) further complicate the connection between microscopic density response and measured dielectric properties \cite{Adler1962,Wiser1963,OnidaReiningRubio2002}.

We present a compact, gauge-invariant bridge that makes predictive optics workable for molecular, crystalline, and heterogeneous materials from RF to UV. Motivated by the need for a unified treatment of interfaces between dissimilar media, we connect chemically specific building blocks to field-level observables across scales. The core idea is to carry the same microscopic inputs—transition dipoles and interaction kernels—from molecules to crystals and heterogeneous media, while treating long-wavelength mixing via a controlled, finite reduction of the dielectric matrix and enforcing length–velocity equality by including the equal-time (diamagnetic/contact) term alongside the paramagnetic current \cite{HybertsenLouie1987,Gajdos2006,GiulianiVignale2005,AversaSipe1995}. Numerically, we emphasize stable workflows based on finite-temperature correlators with controlled analytic continuation and empty-state-free Sternheimer response \cite{VidbergSerene1977,JarrellGubernatis1996,Walker2006,Baroni2001}.

For the reader, this yields a concise, transferable pathway from molecular or per-cell polarizabilities to macroscopic dielectric functions, complex refractive indices, and absorption/heating metrics—so that material changes (functional groups, metal substitution, polymer architecture) can be evaluated directly against observables such as dielectric logging and RF/microwave heating, ellipsometric monitoring of interfacial layers, nanoparticle photothermal response, downhole optics, and complex fluids in porous minerals, under realistic temperature, pressure, and salinity \cite{DresselGruner2002}. We demonstrate the framework with compact validation examples and outline additional use cases as future targets.

\noindent To make this bridge operational—and to outline explicitly the novel ideas— we organize the rest of the paper around a single, end-to-end workflow applied unchanged to molecules, crystals, and interfaces. The steps and section pointers are:

\begin{enumerate}
  \item \textbf{Microscopic inputs.} We start from the same building blocks everywhere—transition dipoles and interaction kernels—computed for molecules or per cell (Sec.~\ref{sec:molecLinResponse}). These are the only ingredients we shall need for the rest of the paper.

  \item \textbf{Finite temperature and spectra.} We handle temperature on the Matsubara axis and, when spectra are required, evaluate directly on the real axis via the complex polarization propagator (Sec.~\ref{sec:matsubaraFormulation}). This keeps numerics stable and units explicit.

  \item \textbf{Periodic systems and local fields.} For crystals and layered/heterogeneous media we work in the $q\!\to\!0$ limit of the dielectric matrix, retaining head/wing/body mixing so local fields are not lost (Sec.~\ref{sec:peridicSystems}). At interfaces, the \emph{application} is done using sheet boundary conditions in the validation section (Sec.~\ref{sec:res-val}, Fig.~\ref{fig:F3}), with the dielectric-matrix background provided by Sec.~\ref{sec:peridicSystems}.

  \item \textbf{Length–velocity bridge (with equal-time term).} We connect per-cell polarizability to conductivity and the dielectric function while explicitly including the equal-time (diamagnetic/contact) term alongside the paramagnetic current—the identity is developed in the periodic $q\!\to\!0$ framework (Sec.~\ref{sec:peridicSystems}) and verified numerically in Sec.~\ref{sec:res-val} (Figs.~\ref{fig:F1}--\ref{fig:F2}).

  \item \textbf{Universal prefactors and units.} We keep prefactors visible via the fine-structure constant $\alpha_{\rm fs}$ so unit consistency is transparent from molecules to 3D solids and 2D sheets (Sec.~\ref{sec:fineStructDependency}). All constants and parameters are listed for reproducibility (Table~\ref{tab:allparams}).

  \item \textbf{Built-in checks.} Before interpreting results, we always confirm the length–velocity overlay of $\varepsilon(\omega)$ (Fig.~\ref{fig:F1}) and saturation of the optical $f$-sum to numerical tolerance (Fig.~\ref{fig:F2}) in Sec.~\ref{sec:res-val}.

  \item \textbf{Observables.} From the same inputs we produce (i) reflectance/absorbance for 2D layers and thin films using the sheet model (Sec.~\ref{sec:res-val}, Fig.~\ref{fig:F3}) and (ii) RF–microwave penetration depths with the expected $\delta\!\propto\!\omega^{-1/2}$ and $\delta\!\propto\!\sigma(T)^{-1/2}$ trends (Sec.~\ref{sec:res-val}, Fig.~\ref{fig:F4}).
\end{enumerate}

Throughout this paper we use atomic units (a.u.): \(e=\hbar=m_e=1\); the speed of light is \(c=1/\alpha_{\rm fs}\), and \(a_0\) is the Bohr radius.

\section{Molecular linear response in the length gauge (finite system)}\label{sec:molecLinResponse}

We shall relate a weak, spatially uniform electric field to the induced molecular dipole, connect the result to spectra and causality, and then show computable forms.

%We use atomic units (a.u.): \(e=\hbar=m_e=1\); the speed of light is \(c=1/\alpha_{\rm fs}\) and \(a_0\) is the Bohr radius.
Throughout this section we adopt the \(e^{+i\omega t}\) Fourier convention and the retarded prescription
\[
\omega \to \omega^{+}\equiv \omega+i0^{+}\quad(\mathrm{Im}\,\omega^{+}>0).
\]
The polarizability can be converted to SI units using
\begin{align}\label{eq:au_si_conversion}
\alpha^{\rm(SI)}(\omega)=4\pi\varepsilon_0 a_0^3\,\alpha^{\rm(a.u.)}(\omega).
\end{align}

In the length gauge, a uniform field couples to the total dipole as\cite{CraigThirunamachandran1998,Mukamel1995,Taghizadeh2017,Taghizadeh2018,Hipolito2018}:
\begin{align}
\delta \hat H(t) = -\,\mathbf E(t)\!\cdot\!\hat{\boldsymbol\mu},\qquad
\hat{\boldsymbol\mu}=-\sum_{i=1}^{N_e}\hat{\mathbf r}_i .
\end{align}
The central response object is the dipole--dipole polarizability tensor.

Linear response gives \cite{Economou2006,FetterWalecka2018}
\begin{align}\label{eq:polarizability_correlator}
\alpha_{ij}(\omega)\equiv\chi^R_{\mu_i\mu_j}(\omega)
=\int_0^\infty\!dt\,e^{i\omega^{+}t}\,(-i)\,\big\langle[\hat\mu_i(t),\hat\mu_j(0)]\big\rangle_0,
\end{align}
with thermal average $\braket{\, \cdot\, }_0 \equiv \Tr(\rho_0\,\cdot)$ in \(\hat\rho_0=e^{-\beta \hat H_0}/\mathcal Z\), \(\beta=1/k_BT\), \(\mathcal Z=\sum_n e^{-\beta E_n}\).
Causality implies \(\alpha_{ij}(-\omega)=\alpha_{ji}(\omega)^\ast\) and \(\Im\,\alpha_{ii}(\omega)\ge 0\) for \(\omega>0\).

Inserting a full set of eigenstates of the free Hamiltonian \(\hat H_0|n\rangle=E_n|n\rangle\) with \(\omega_{nm}=E_n-E_m\) and \(\mu^i_{mn}=\langle m|\hat\mu_i|n\rangle\) we find at finite \(T\) \cite{Economou2006,FetterWalecka2018}
\begin{align}
\alpha_{ij}(\omega)
=\frac{1}{\mathcal Z}\sum_{m,n}\!\big(e^{-\beta E_m}-e^{-\beta E_n}\big)\,
\frac{\mu^i_{mn}\mu^j_{nm}}{\omega^{+}-\omega_{nm}} .
\end{align}
Taking \(T\!\to\!0\) and combining upward/downward transitions yields
\begin{align}\label{eq:polarizability_T0}
\alpha_{ij}(\omega)
=2\sum_{n\neq0}\frac{\omega_{n0}\,\mu^i_{0n}\mu^j_{n0}}{\omega_{n0}^2-(\omega^{+})^2},
\end{align}
with static limit \(\alpha_{ij}(0)=2\sum_{n\neq0}\mu^i_{0n}\mu^j_{n0}/\omega_{n0}\).

For randomly oriented samples,
\begin{align}
\alpha_{\rm iso}(\omega)=\tfrac{1}{3}\,\mathrm{Tr}\,\boldsymbol\alpha(\omega)
=\frac{2}{3}\sum_{n\neq0}\frac{\omega_{n0}\,|\boldsymbol\mu_{0n}|^2}{\omega_{n0}^2-(\omega^{+})^2}.
\end{align}
Hence
\begin{align}
\Im\,\alpha_{\rm iso}(\omega)=\frac{\pi}{3}\sum_{n\neq0}
|\boldsymbol\mu_{0n}|^2\Big[\delta(\omega-\omega_{n0})-\delta(\omega+\omega_{n0})\Big],
\end{align}
and defining \(f_{0n}=\tfrac{2}{3}\,\omega_{n0}|\boldsymbol\mu_{0n}|^2\) gives the Thomas–Reiche–Kuhn (TRK) sum rule \cite{OnidaReiningRubio2002}:
\begin{align}
\sum_{n}f_{0n}=N_e
\quad\Longleftrightarrow\quad
\frac{2}{\pi}\int_0^\infty\! d\omega\,\omega\,\Im\,\alpha_{\rm iso}(\omega)=\frac{N_e}{3}.
\end{align}

The shift \(\omega\!\to\!\omega^{+}\) enforces retardation. Using the Sokhotski--Plemelj identity,
\(1/(x\!\pm\! i0^{+})=\mathcal P(1/x)\mp i\pi\delta(x)\), where \(\mathcal P\) denotes the
Cauchy principal value, one obtains the Kramers--Kronig (KK) relations \cite{Toll1956,Economou2006}:
\begin{align}
\Re\,\alpha_{ij}(\omega)&=\frac{2}{\pi}\,\mathcal P\!\int_0^\infty d\omega'\,
\frac{\omega'\,\Im\,\alpha_{ij}(\omega')}{\omega'^2-\omega^2}, \\
\Im\,\alpha_{ij}(\omega)&=-\frac{2\omega}{\pi}\,\mathcal P\!\int_0^\infty d\omega'\,
\frac{\Re\,\alpha_{ij}(\omega')}{\omega'^2-\omega^2}.
\end{align}
In practice, the infinitesimal shift \(0^+\) should be replaced by a small broadening \(\eta>0\) or evaluated using a discrete Hilbert transform with careful tail treatment. In a HF/KS one--electron basis \(\{\phi_p\}\) following a Fermi-Dirac distribution with weights $f_p$ with creation/annihilation operators \(a_p^\dagger,a_q\),
\begin{align}
\hat\mu_i=\sum_{pq}\mu^{(i)}_{pq}\,a_p^\dagger a_q,\qquad
\mu^{(i)}_{pq}=\int \phi_p^\ast(\mathbf r)\,(-r_i)\,\phi_q(\mathbf r)\,d\mathbf r.
\end{align}
The independent--particle susceptibility is
\begin{align}
\chi^{0}_{ij}(\omega)=2\sum_{pq}(f_p-f_q)\,\mu^{(i)}_{pq}\mu^{(j)}_{qp}\,
\frac{1}{\omega+i\eta-(\varepsilon_q-\varepsilon_p)} .
\end{align}
For a closed shell at \(T=0\),
\begin{align}
\chi^{0}_{ij}(\omega)=2\sum_{ka}\mu^{(i)}_{ka}\mu^{(j)}_{ak}
\left[\frac{1}{\omega+i\eta-\Delta\varepsilon_{ka}}
-\frac{1}{\omega+i\eta+\Delta\varepsilon_{ka}}\right],
\end{align}
with $\Delta\varepsilon_{ka}=\varepsilon_a-\varepsilon_k$ .

Electron interactions dress \(\chi^0\). In particle--hole space \cite{PetersilkaGossmannGross1996,Casida1995,OnidaReiningRubio2002,Helgaker2000Online2014}
\begin{align}
\chi(\omega)=\chi^0(\omega)+\chi^0(\omega)\,K(\omega)\,\chi(\omega),
\end{align}
with kernel
\begin{align}
K_{ka,lb}(\omega)&=(ka|v|lb)+(ka|f_{\rm xc}(\omega)|lb)+K^{x,{\rm nonloc}}_{ka,lb}, \\
v(\mathbf r,\mathbf r')&=\frac{1}{|\mathbf r-\mathbf r'|},
\end{align}

where we used the convention for indices as Cartesian axes \(i,j\); occupied orbitals \(k,l\); virtual orbitals \(a,b\).\\
We shall now represent this equation as a linear system, which makes treatment more straightforward.\\ 
(i) \emph{Frequency--domain (CPP/CPKS)} \cite{Helgaker2000Online2014}:
\begin{align}
\begin{pmatrix}
\mathbf A-\omega\mathbf 1 & \mathbf B\\
\mathbf B & \mathbf A+\omega\mathbf 1
\end{pmatrix}
\begin{pmatrix}\mathbf X^{(j)}\\ \mathbf Y^{(j)}\end{pmatrix}
=-
\begin{pmatrix}\boldsymbol\mu^{(j)}\\ \boldsymbol\mu^{(j)}\end{pmatrix},
\end{align}
with \(A_{ka,lb}=\Delta\varepsilon_{ka}\,\delta_{kl}\delta_{ab}+K_{ka,lb}\),
\(B_{ka,lb}=K_{kb,la}\), \(\mu^{(j)}_{ka}=\langle k|(-r_j)|a\rangle\), and
\begin{align}
\alpha_{ij}(\omega)=\sum_{ka}\mu^{(i)}_{ak}\,[X^{(j)}_{ka}(\omega)+Y^{(j)}_{ka}(\omega)].
\end{align}
(ii) \emph{Excitation--mode (Casida)} \cite{Casida1995,OnidaReiningRubio2002}:
\begin{align}
\begin{pmatrix}\mathbf A&\mathbf B\\ \mathbf B&\mathbf A\end{pmatrix}
\begin{pmatrix}\mathbf X^{S}\\ \mathbf Y^{S}\end{pmatrix}
=\omega_S
\begin{pmatrix}\mathbf 1&\mathbf 0\\ \mathbf 0&-\mathbf 1\end{pmatrix}
\begin{pmatrix}\mathbf X^{S}\\ \mathbf Y^{S}\end{pmatrix}
\ \Rightarrow\
\boldsymbol\Omega\,\mathbf F^{S}=\omega_S^2\,\mathbf F^{S},
\end{align}
with
\begin{align}
\Omega_{ka,lb}=\delta_{kl}\delta_{ab}\,\Delta\varepsilon_{ka}^2
+4\sqrt{\Delta\varepsilon_{ka}}\,K_{ka,lb}\,\sqrt{\Delta\varepsilon_{lb}}.
\end{align}
Defining transition dipoles \(t^{(i)}_{S}=\sum_{ka}(X^{S}_{ka}+Y^{S}_{ka})\,\mu^{(i)}_{ak}\) recovers
\begin{align}
\alpha_{ij}(\omega)=\sum_{S}\frac{2\,\omega_S\,t^{(i)}_{S}t^{(j)}_{S}}{\omega_S^{2}-(\omega+i\eta)^{2}}.
\end{align}

Length- and velocity-gauge forms are equivalent in a complete basis; any difference reflects basis truncation \cite{CraigThirunamachandran1998,Mukamel1995,Taghizadeh2018}.
Besides \(\alpha_{\rm iso}\), it is useful to report
\(\Delta\alpha(\omega)=\sqrt{\tfrac{3}{2}\,\alpha'_{ij}\alpha'_{ij}}\) with
\(\alpha'_{ij}=\alpha_{ij}-\tfrac{1}{3}\delta_{ij}\mathrm{Tr}\,\boldsymbol\alpha\).
The extinction (absorption) cross section in the dipole approximation is \cite{Economou2006}
\begin{align}
\sigma_{\rm abs}(\omega)
=\frac{\omega}{\varepsilon_0 c}\,\Im\,\alpha_{\rm iso}^{(\mathrm{SI})}(\omega)
=4\pi\,\alpha_{\rm fs}\,\omega\,a_0^2\,\Im\,\alpha_{\rm iso}^{(\mathrm{a.u.})}(\omega).
\end{align}

\section{Finite-temperature Matsubara formulation and analytic continuation}\label{sec:matsubaraFormulation}

%We recall that the present work uses atomic units (\(e=\hbar=m_e=1\)) and \(\beta=1/k_B T\).
We recall that unperturbed eigenstates satisfy \(\hat H_0|n\rangle=E_n|n\rangle\) and follow the partition function \(\mathcal Z=\sum_n e^{-\beta E_n}\), with \(\beta=1/k_B T\).
The total dipole operator is \(\hat{\boldsymbol\mu}=-\sum_{i=1}^{N_e}\hat{\mathbf r}_i\) in a.u.\ (in mixed units: \(-e\sum_i\hat{\mathbf r}_i\)).
At finite \(T\) we use fluctuations \(\delta\hat\mu_i=\hat\mu_i-\langle\hat\mu_i\rangle\) and define the imaginary-time (Matsubara) dipole–dipole susceptibility \cite{Mahan2000,FetterWalecka2018,Economou2006}:
\begin{align}
    \chi_{ij}(\tau)&\equiv
\big\langle T_\tau\,\delta\hat\mu_i(\tau)\,\delta\hat\mu_j(0)\big\rangle_0,\qquad
0\le \tau<\beta,\\
\hat O(\tau)&=e^{\tau \hat H_0}\hat O e^{-\tau \hat H_0},
\end{align}
where $T_\tau$ is the imaginary-time ordering operator and $\braket{\cdot}_0 = \Tr(\rho_0 \cdot)$.
Bosonic KMS periodicity holds,
\begin{align}
    \chi_{ij}(\tau+\beta)=\chi_{ij}(\tau),
\end{align}
so we expand in bosonic Matsubara frequencies
\begin{align}
    \omega_m=\frac{2\pi m}{\beta},\qquad m\in\mathbb Z,
\end{align}
with Fourier transform
\begin{align}
    \chi_{ij}(i\omega_m)=\int_0^\beta d\tau\,e^{-i(i\omega_m) \tau}\,\chi_{ij}(\tau).
\end{align}

Inserting a complete set of \(|n\rangle\) (Lehmann representation) with \(\omega_{nm}=E_n-E_m\) and \(\mu^i_{mn}=\langle m|\hat\mu_i|n\rangle\) gives \cite{Mahan2000,FetterWalecka2018,Economou2006}
\begin{align}
    \chi_{ij}(i\omega_m)=\frac{1}{\mathcal Z}\sum_{m,n}
\frac{e^{-\beta E_m}-e^{-\beta E_n}}{i\omega_m-\omega_{nm}}\;
\mu^i_{mn}\,\mu^j_{nm}.
\end{align}
Equivalently, introducing the (diagonal-positive) spectral function
\begin{align}
    S_{ij}(\omega)\equiv \frac{1}{\mathcal Z}\sum_{m,n}e^{-\beta E_m}\,
    \mu^i_{mn}\mu^j_{nm}\,\delta(\omega-\omega_{nm}),
\end{align}
one has detailed balance \(S_{ji}(-\omega)=e^{-\beta\omega}S_{ij}(\omega)\) and the dispersion representations
\begin{align}
    \chi_{ij}(i\omega_m)&=\int_{-\infty}^{+\infty}\!d\omega'\,
    \frac{S_{ij}(\omega')-S_{ji}(-\omega')}{i\omega_m-\omega'},\\
    \chi^{R}_{ij}(\omega)&=\int_{-\infty}^{+\infty}\!d\omega'\,
    \frac{S_{ij}(\omega')-S_{ji}(-\omega')}{\omega-\omega'+i0^+}.
\end{align}
These justify the analytic-continuation rule
\begin{align}
    \chi^{R}_{ij}(\omega)&=\chi_{ij}(i\omega_m\!\to\!\omega+i0^+), \\
\alpha_{ij}(\omega)&\equiv \chi^{R}_{\mu_i\mu_j}(\omega),
\end{align}
and yield the fluctuation–dissipation theorem \cite{Mahan2000,FetterWalecka2018}
\begin{align}
    \Im\,\chi^{R}_{ij}(\omega)=\pi\big(1-e^{-\beta\omega}\big)\,S_{ij}(\omega),
\end{align}
so for \(i=j\) one has \(\Im\,\alpha_{ii}(\omega)\ge 0\) for \(\omega>0\).

In practice, we often begin from an independent-particle (IP) reference with one-electron orbitals \(\{\phi_p\}\), with associated energies \(\varepsilon_p\), and following a Fermi distribution with weights/factors \(f_p=[e^{\beta(\varepsilon_p-\mu)}+1]^{-1}\), with $\mu$ the chemical potential.
The IP dipole–dipole susceptibility on the Matsubara axis is
\begin{align}
    \chi^{0}_{ij}(i\omega_m)&=
2\sum_{p,q}\frac{f_p-f_q}{i\omega_m-(\varepsilon_q-\varepsilon_p)}\;
\mu^{(i)}_{pq}\,\mu^{(j)}_{qp}, \\
\mu^{(i)}_{pq}&=\langle \phi_p|(-r_i)|\phi_q\rangle,
\end{align}
where the factor 2 accounts for spin degeneracy ( and should be omitted for explicit spinor bases).
Interactions beyond IP are included via the Dyson equation in particle–hole space\cite{Economou2006},
\begin{align}
\chi(i\omega_m)=\chi^0(i\omega_m)+\chi^0(i\omega_m)\,K(i\omega_m)\,\chi(i\omega_m),
\end{align}
with kernel \(K_{ia,jb}=(ia|v|jb)+(ia|f_{\rm xc}|jb)+K^{x,{\rm nonloc}}_{ia,jb}\).
For adiabatic local density approximation (ALDA) kernels and many hybrid density functionals, \(K\) is \(\omega\)-independent.

If \(K\) is frequency independent, the static polarizability is available \emph{on the imaginary axis} without continuation\cite{Economou2006}:
\begin{align}
    \alpha_{ij}(0)=\chi_{ij}(i\omega_m{=}0)
=\big[\mathbf 1-\chi^0(0)\,K\big]^{-1}\chi^0_{ij}(0).
\end{align}
For \(\omega\neq 0\), one must obtain \(\chi^{R}_{ij}(\omega)\) from \(\chi_{ij}(i\omega_m)\).
Because analytic continuation from discrete Matsubara data is ill-posed, common strategies are:
Pad{\'e}/rational approximants \cite{VidbergSerene1977}, maximum-entropy/Bayesian reconstruction of \(S_{ij}(\omega)\) followed by KK \cite{JarrellGubernatis1996}, or bypassing continuation entirely by a direct real-axis solver such as the complex polarization propagator (evaluate at \(\omega\!\to\!\omega+i\eta\)) \cite{Helgaker2000Online2014,Walker2006}:
\begin{align}
    \chi^{R}_{ij}(\omega)\approx \chi_{ij}(\omega+i\eta),\qquad \eta>0.
\end{align}

As finite-\(T\) checks: (i) Kubo-Martin-Schwinger\cite{Kubo1957,MartinSchwinger1959} (KMS) periodicity \(\chi_{ij}(\tau+\beta)=\chi_{ij}(\tau)\) holds; (ii) \(\chi^{R}_{ij}(\omega)\) is analytic in the upper half-plane and obeys \(\chi^{R}_{ij}(-\omega)=\chi^{R\ast}_{ji}(\omega)\); (iii) the longitudinal \(f\)-sum rule keeps its temperature-independent form for the isotropic polarizability (TRK) \cite{OnidaReiningRubio2002,FetterWalecka2018}:
\begin{align}
\frac{2}{\pi}\int_0^\infty d\omega\,\omega\,\Im\,\alpha_{\rm iso}(\omega)=\frac{N_e}{3}.
\end{align}

\section{Periodic boundary conditions (length gauge, \texorpdfstring{$q\!\to\!0$}{q->0})}
\label{sec:peridicSystems}

A strictly uniform longitudinal field cannot be represented by a periodic scalar potential. In the \emph{length gauge}, we excite with a long-wavelength \emph{longitudinal} scalar potential and take $q\!\to\!0$ (choosing $\mathbf E\parallel\mathbf q$):
\begin{align}
    \delta \phi_{\mathbf q}(\omega) \;=\; i\,\frac{\mathbf q\!\cdot\!\mathbf E(\omega)}{q^2}, \qquad  q\to 0,
\end{align}
so the microscopic perturbation is
\begin{align}
    \delta \hat V \;=\; \int d\mathbf r\, \hat n(\mathbf r)\,\delta\phi_{\mathbf q}(\omega)\,e^{i\mathbf q\cdot\mathbf r}.
\end{align}
The relevant response object is the density–density polarizability (see Adler \& Wiser) \cite{Adler1962,Wiser1963,OnidaReiningRubio2002}
\begin{align}
    \chi^0_{GG'}(\mathbf q,\omega) & = \frac{2}{\Omega}
\sum_{n m}\!\int_{\rm BZ}\!\frac{d^3k}{(2\pi)^3}\,
\frac{f_{n\mathbf k}-f_{m,\mathbf k+\mathbf q}}{\hbar\omega + \varepsilon_{n\mathbf k} - \varepsilon_{m,\mathbf k+\mathbf q}+i0^+} \nonumber\\
&\times\langle n\mathbf k|e^{-i(\mathbf q+\mathbf G)\cdot \mathbf r}|m,\mathbf k+\mathbf q\rangle\,
\langle m,\mathbf k+\mathbf q|e^{+i(\mathbf q+\mathbf G')\cdot \mathbf r}|n\mathbf k\rangle,
\end{align}
where $\Omega$ is the volume of the first Brillouin Zone (BZ) (cell), and the polarizability is dressed by Coulomb and an xc kernel through
\begin{align}
\chi&=\chi^0 + \chi^0\,(v+f_{\rm xc})\,\chi, \\
v_{GG'}(\mathbf q)&=\frac{4\pi}{|\mathbf q+\mathbf G|^2}\,\delta_{GG'}\quad\text{(Hartree a.u.)}.
\end{align}
The macroscopic longitudinal dielectric function follows from the inverse head \cite{HybertsenLouie1987,Gajdos2006,OnidaReiningRubio2002}:
\begin{align}
    \varepsilon_M(\omega)&=\lim_{\mathbf q\to 0} \frac{1}{\varepsilon^{-1}_{00}(\mathbf q,\omega)},\\
    \varepsilon^{-1}&=1+v\chi,\ \ \varepsilon=1-v\chi.
\end{align}

To expose the small-$q$ structure (head/wing/body) and inter/intraband content, we  split the response function different regions
\begin{align}
\chi^0_{GG'}=\chi^{0,{\rm inter}}_{GG'}+\chi^{0,{\rm intra}}_{GG'}.
\end{align}
Writing $|n\mathbf k\rangle=e^{i\mathbf k\cdot\mathbf r}|u_{n\mathbf k}\rangle$ and defining the \emph{charge vertices}
\begin{align}
    \Gamma^{G}_{nm}(\mathbf k;\mathbf q)&\equiv
\langle n\mathbf k|e^{-i(\mathbf q+\mathbf G)\cdot \mathbf r}|m,\mathbf k+\mathbf q\rangle \nonumber \\
&=\langle u_{n\mathbf k}|e^{-i\mathbf G\cdot\mathbf r}\,e^{-i\mathbf q\cdot\mathbf r}|u_{m,\mathbf k+\mathbf q}\rangle.
\end{align}
With $|u_{m,\mathbf k+\mathbf q}\rangle=|u_{m\mathbf k}\rangle+(\mathbf q\!\cdot\!\nabla_{\mathbf k})|u_{m\mathbf k}\rangle+\mathcal O(q^2)$ and $e^{-i\mathbf q\cdot\mathbf r}=1-i\,\mathbf q\!\cdot\!\mathbf r+\mathcal O(q^2)$, the covariant position elements \cite{Resta1994,AversaSipe1995,Ventura2017}
\begin{align}
    r^{i}_{nm}(\mathbf k)&=i\langle u_{n\mathbf k}|\partial_{k_i}u_{m\mathbf k}\rangle\ (n\neq m), \\
    A^{i}_{n}(\mathbf k)&=i\langle u_{n\mathbf k}|\partial_{k_i}u_{n\mathbf k}\rangle,
\end{align}
give
\begin{align}
\Gamma^{0}_{nm}(\mathbf k;\mathbf q)
&=\delta_{nm}-i\,\mathbf q\!\cdot\!\Big[\delta_{nm}\,\mathbf A_n(\mathbf k)-\mathbf r_{nm}(\mathbf k)\Big]+\mathcal O(q^2),\\
\Gamma^{G\neq 0}_{nm}(\mathbf k;\mathbf q)
&= S^{G}_{nm}(\mathbf k) + \mathcal O(q), 
\end{align}
with 
$S^{G}_{nm}(\mathbf k)\equiv\langle u_{n\mathbf k}|e^{-i\mathbf G\cdot\mathbf r}|u_{m\mathbf k}\rangle$.

Occupations expand as
\begin{align}
f_{n\mathbf k}-f_{m,\mathbf k+\mathbf q}=
\begin{cases}
f_{n\mathbf k}-f_{m\mathbf k}+\mathcal O(q), & n\neq m \ (\text{interband}),\\[2pt]
-\mathbf q\!\cdot\!(\nabla_{\mathbf k} f_{n\mathbf k})+\mathcal O(q^2), & n=m \ (\text{intraband}).
\end{cases}
\end{align}

For the \emph{Head (with $G=G'=0$).} In the interband up to leading order, the susceptibility becomes
\begin{align}
\chi^{0,{\rm inter}}_{00}(\mathbf q,\omega)
= &\frac{2}{\Omega}\sum_{n\neq m}\!\int_{\rm BZ}\!\frac{d^3k}{(2\pi)^3}\,
\frac{f_{n\mathbf k}-f_{m\mathbf k}}{\hbar\omega+\varepsilon_{n\mathbf k}-\varepsilon_{m\mathbf k}+i0^+}\, \nonumber \\
&\big(q_i r^{i}_{nm}\big)\big(q_j r^{j}_{mn}\big)+\mathcal O(q^3),
\end{align}
with Einstein implicit summation on $i,j$. Note that the leading order is then $\mathcal O(q^2)$. In the intraband, we have
\begin{align}\label{eq:head_intraband}
\chi^{0,{\rm intra}}_{00}(\mathbf q,\omega)
=\frac{2}{\Omega}\sum_{n}\!\int_{\rm BZ}\!\frac{d^3k}{(2\pi)^3}\,
\frac{-\,\mathbf q\!\cdot\!(\nabla_{\mathbf k} f_{n\mathbf k})}{\omega-\mathbf q\!\cdot\!\mathbf v_{n\mathbf k}+i0^+}
\ \Big[1+\mathcal O(q)\Big],
\end{align}
where we recall that $\mathbf v_{n\mathbf k} \equiv [\mathbf{D}_{\mathbf k}, \hat{H}_0]_{n}/\hbar$ where $\mathbf{D}_{\mathbf k}$ is the covariant derivative.
Expanding equation \ref{eq:head_intraband} at fixed $\omega\neq 0$, yields
\begin{align}
\chi^{0,{\rm intra}}_{00}(\mathbf q,\omega)
\simeq &\frac{q_i q_j}{\omega(\omega+i0^+)}\;
\frac{2}{\Omega}\sum_{n}\!\int_{\rm BZ}\!\frac{d^3k}{(2\pi)^3}\, \nonumber \\
& \Big(-\frac{\partial f_{n\mathbf k}}{\partial \varepsilon_{n\mathbf k}}\Big)\,v^i_{n}(\mathbf k)\,v^j_{n}(\mathbf k).
\end{align}
Thus, for finite $\omega$ the head vanishes as $q^2$ (finite macroscopic permittivity). In metals, the $\omega\!\to\!0$ limit requires Drude/hydrodynamic resummation (see below).

In the \emph{Wings ($G=0$, $G'\neq 0$ or vice versa).} The interband is:
\begin{align}
\chi^{0,{\rm inter}}_{0G'}(\mathbf q,\omega)
= &\frac{2}{\Omega}\sum_{n\neq m}\!\int_{\rm BZ}\!\frac{d^3k}{(2\pi)^3}\,
\frac{f_{n\mathbf k}-f_{m\mathbf k}}{\hbar\omega+\varepsilon_{n\mathbf k}-\varepsilon_{m\mathbf k}+i0^+}\, \nonumber \\
&\big(q_i r^{i}_{nm}\big)\,S^{G'}_{mn}(\mathbf k)+\mathcal O(q^2),
\end{align}
i.e.\ $\chi^{0,{\rm inter}}_{0G'}(\mathbf q,\omega) \sim \mathcal O(q)$. Wings mediate \emph{local-field effects} by coupling the macroscopic sector to microscopic harmonics $G'\!\neq\!0$; intraband wings behave similarly.

For the \emph{Body ($G\neq 0$, $G'\neq 0$)}, the interband dominates at $\mathcal O(1)$:
\begin{align}
\chi^{0,{\rm inter}}_{GG'}(\mathbf q,\omega)
= &\frac{2}{\Omega}\sum_{n\neq m}\!\int_{\rm BZ}\!\frac{d^3k}{(2\pi)^3}\,
\frac{f_{n\mathbf k}-f_{m\mathbf k}}{\hbar\omega+\varepsilon_{n\mathbf k}-\varepsilon_{m\mathbf k}+i0^+}\, \nonumber \\
& S^{G}_{nm}(\mathbf k)\,S^{G'}_{mn}(\mathbf k)+\mathcal O(q),
\end{align}
while intraband body terms are subleading ($\propto q$). The Coulomb metric $v_{GG'}\!\propto\!|\mathbf q+\mathbf G|^{-2}\delta_{GG'}$ and any $f_{\rm xc}$ mix head/wing/body, so the correct $\varepsilon_M$ requires the inverse head $\varepsilon^{-1}_{00}$ to include wing/body  into the sum \cite{HybertsenLouie1987,OnidaReiningRubio2002}.

\emph{Macroscopic ($q=0$) length-gauge form.} With polarization $\mathbf P=\langle \hat{\boldsymbol\mu}\rangle/\Omega$ and, in SI,
\begin{align}
    \chi_{e,ij}(\omega)&=\frac{\alpha^{\rm cell}_{ij}(\omega)}{\varepsilon_0\,\Omega}, \\
    \varepsilon_{r,ij}(\omega)&=\delta_{ij}+\chi_{e,ij}(\omega),
\end{align}
the $q\!\to\!0$ interband/intraband contributions can be written in the Bloch length gauge \cite{AversaSipe1995,Resta1994,Ventura2017} as
\begin{align}
\chi^{\rm inter}_{ij}(\omega)=&\frac{e^2}{\hbar\,\Omega}
\sum_{n\neq m}\!\int_{\rm BZ}\!\frac{d^3k}{(2\pi)^3}
\frac{f_{n\mathbf k}-f_{m\mathbf k}}{\omega_{mn}(\mathbf k)} \nonumber \\
&\frac{r^{i}_{nm}(\mathbf k)\,r^{j}_{mn}(\mathbf k)}{\omega_{mn}(\mathbf k)-\omega-i\eta}
+{\rm c.c.},\\
\chi^{\rm intra}_{ij}(\omega)=&
-\frac{e^2}{\Omega}\sum_{n}\!\int_{\rm BZ}\!\frac{d^3k}{(2\pi)^3}
\frac{\partial f_{n\mathbf k}}{\partial \varepsilon_{n\mathbf k}}\, \nonumber \\
&\frac{v^{i}_n(\mathbf k)\,v^{j}_n(\mathbf k)}{\omega(\omega+i\gamma)},
\end{align}
with diamagnetic weight
\begin{align}
    D_{ij}=\frac{e^2}{\Omega}\sum_{n}\!\int_{\rm BZ}\!\frac{d^3k}{(2\pi)^3}
\Big(-\frac{\partial f_{n\mathbf k}}{\partial \varepsilon_{n\mathbf k}}\Big)
\,[m^{-1}_{n}(\mathbf k)]_{ij},
\end{align}
with $m$ the effective mass tensor.
Neglecting local fields (wings/body), the head directly gives (Gaussian/au) \cite{OnidaReiningRubio2002}
\begin{align}
    \varepsilon_M(\omega)=1-\lim_{\mathbf q\to 0}\frac{4\pi e^2}{q^2}\,\chi_{00}(\mathbf q,\omega).
\end{align}

\begin{align}
  \boldsymbol\sigma(\omega) &= -\,i\,\varepsilon_0\,\omega\,\boldsymbol\chi_e(\omega)
  \;=\; -\,\frac{i\omega}{\Omega}\,\boldsymbol\alpha^{\rm cell}(\omega),\\
  \boldsymbol\varepsilon_r(\omega) &= \mathbf 1 + \boldsymbol\chi_e(\omega)
  \;=\; \mathbf 1 + \frac{i}{\varepsilon_0\omega}\,\boldsymbol\sigma(\omega).
\end{align}
and, with the current–current Kubo form \cite{Kubo1957,MartinSchwinger1959,Greenwood1958,GiulianiVignale2005}

\begin{align}
    \sigma_{ij}(\omega)&= \frac{1}{i(\omega+i0^+)}\Big[\Pi_{ij}^R(\omega)+D_{ij}\Big], \\
    \Pi_{ij}^R(\omega)&= -\frac{i}{\hbar}\!\int_0^\infty\!dt\, e^{i(\omega+i0^+)t}\,
\langle[\,J_{p,i}(t),J_{p,j}(0)\,]\rangle,
\end{align}
one finds the exact identity
\begin{align}
    \Pi_{ij}^R(\omega)=\frac{\omega^2}{\Omega}\,\alpha^{\rm cell}_{ij}(\omega) - D_{ij}
\ \Rightarrow\
\sigma_{ij}(\omega) = -\frac{i\omega}{\Omega}\,\alpha^{\rm cell}_{ij}(\omega),
\end{align}
i.e.\ length and velocity gauges are identical in a complete basis. Modern tight-binding analyses reach the same conclusion once the diamagnetic term is treated correctly \cite{Schuler2021,Taghizadeh2017,Taghizadeh2018,Ventura2017}.

\emph{Metallic $q\!\to\!0$ vs.\ $\omega\!\to\!0$ (Drude/hydrodynamic resummation).}
In metals the intraband head yields a non-commuting limit. A controlled interpolation between finite-$\omega$ Drude behavior and static Thomas–Fermi screening is \cite{DresselGruner2002,GiulianiVignale2005}
\begin{align}
    \varepsilon_L(\omega,q)&= 1 - \frac{\omega_p^2}{\omega(\omega+i\gamma)-\beta^2 q^2},\\
    \beta^2 &\simeq \tfrac{3}{5}v_F^2, \\
\omega_p^2 &=
\begin{cases}
\displaystyle 4\pi n e^2/m,& \text{(Gaussian/au)},\\[2pt]
\displaystyle n e^2/(\varepsilon_0 m),& \text{(SI)},
\end{cases}
\end{align}
equivalently for the microscopic head
\begin{align}
    \chi_{00}(q,\omega)\simeq \frac{n e^2}{m}\;
    \frac{q^2}{\omega(\omega+i\gamma)-\beta^2 q^2},
\end{align}
so that $\varepsilon_M(\omega)=1-\lim_{q\to 0}[4\pi e^2/q^2]\chi_{00}(q,\omega)$ reproduces Drude at $q\rightarrow0$ and, at $\omega{=}0$, yields $\varepsilon_L(0,q)=1+k_{\rm TF}^2/q^2$ with $k_{\rm TF}^2=\omega_p^2/\beta^2$.

\section{Dependence on \texorpdfstring{$\alpha_{\mathrm{fs}}$}{alpha\_fs}: length-gauge polarizability to Kubo conductivity}
\label{sec:fineStructDependency}

We retain the $e^{+i\omega t}$ time dependence used in previous sections, so that
$\mathbf J(\omega)=i\omega\,\mathbf P(\omega)$ and
$\boldsymbol\sigma(\omega)=i\varepsilon_0\omega\,\boldsymbol\chi_e(\omega)$,
hence $\varepsilon_r(\omega)=1-\tfrac{i}{\varepsilon_0\omega}\,\boldsymbol\sigma(\omega)$ \cite{DresselGruner2002}.
In this section $\alpha_{\rm fs}$ denotes the \emph{fine-structure constant}, whereas $\boldsymbol{\alpha}(\omega)$ denotes the \emph{molecular (or per-cell) polarizability tensor}. 
Our goal is to render explicitly how universal prefactors built from $e$, $\hbar$, $c$, and $\varepsilon_0$ can be rewritten in terms of $\alpha_{\rm fs}$ so that the overall scale of response functions becomes transparent, while the material-specific physics remains in (nearly) dimensionless correlation kernels.

\begin{align}
\alpha_{\rm fs} &= \frac{e^2}{4\pi\varepsilon_0\hbar c},\\
\frac{e^2}{\hbar} &= 4\pi\varepsilon_0 c\,\alpha_{\rm fs},\\
\frac{e^2}{h} &= 2\,\varepsilon_0 c\,\alpha_{\rm fs}.
\end{align}

Within linear response, the Kubo conductivity factors into a universal prefactor and a dimensionless correlator,
\begin{align}
\sigma_{ij}(\omega) = \frac{e^2}{\hbar}\,\mathcal K_{ij}(\omega)
= 4\pi\varepsilon_0 c\,\alpha_{\rm fs}\,\mathcal K_{ij}(\omega),
\end{align}
and its relation to the electric susceptibility and dielectric function is
\begin{align}
\sigma_{ij}(\omega) &= -\,i\,\varepsilon_0\omega\,\chi_{e,ij}(\omega),\\
\varepsilon_{r,ij}(\omega) &= \delta_{ij}+\chi_{e,ij}(\omega)
= \delta_{ij} + \frac{i}{\varepsilon_0\omega}\,\sigma_{ij}(\omega),
\end{align}
as in standard Kubo/linear-response treatments \cite{Kubo1956,Kubo1957,GiulianiVignale2005,DresselGruner2002}.

At the molecular level, the isotropic absorption cross section in atomic units shows the linear dependence on $\alpha_{\rm fs}$:
\begin{align}
\sigma_{\rm abs}^{\rm mol}(\omega)
&= 4\pi\,\alpha_{\rm fs}\,\omega\,\Im\,\alpha_{\rm iso}^{\rm(a.u.)}(\omega)\,a_0^2, \\
\alpha_{\rm iso}(\omega)&=\tfrac{1}{3}\,\mathrm{Tr}\,\boldsymbol{\alpha}(\omega),
\end{align}
with $a_0$ the Bohr radius and, in SI units, one converts the polarizability and writes
\begin{align}
\alpha^{\rm(SI)}(\omega) &= 4\pi\varepsilon_0 a_0^3\,\alpha^{\rm(a.u.)}(\omega),\\
\sigma_{\rm abs}^{\rm mol}(\omega)
&= \frac{\omega}{\varepsilon_0 c}\,\Im\,\alpha_{\rm iso}^{\rm(SI)}(\omega),
\end{align}
see, e.g., Craig \textit{et. al} (1998) or Mukamel (1995)\cite{CraigThirunamachandran1998,Mukamel1995}.

For periodic or dense systems, connect the \emph{per-cell} dipole response to macroscopic quantities (cell volume $\Omega$):
\begin{align}
\chi_{e,ij}(\omega) &= \frac{\alpha^{\rm cell}_{ij}(\omega)}{\varepsilon_0\,\Omega},\\
\varepsilon_{r,ij}(\omega) &= \delta_{ij} + \chi_{e,ij}(\omega),\\
\sigma_{ij}(\omega) &= -\,\frac{i\omega}{\Omega}\,\alpha^{\rm cell}_{ij}(\omega),
\end{align}
consistent with microscopic--macroscopic mappings in solids \cite{OnidaReiningRubio2002,HybertsenLouie1987,Gajdos2006}.

For free carriers (Drude metals) the spectral weight is set by the plasma frequency,
\begin{align}
\omega_p^2
= \frac{n e^2}{\varepsilon_0 m^\ast}
= 4\pi\,\alpha_{\rm fs}\,c\,\frac{n\hbar}{m^\ast},
\end{align}
where $n$ is the carrier density and $m^\ast$ is the (band/optical) effective mass.
% The second equality uses $\alpha_{\rm fs}=e^2/(4\pi\varepsilon_0\hbar c)$.

In strictly two dimensions the natural prefactor is $e^2/h$, hence
\begin{align}
\frac{e^2}{h} = 2\,\varepsilon_0 c\,\alpha_{\rm fs}.
\end{align}
In the thin-sheet, normal-incidence limit in vacuum ($|Z_0\sigma_{\rm 2D}|\ll 1$, $Z_0=\sqrt{\mu_0/\varepsilon_0}=1/(\varepsilon_0 c)$),
\begin{align}
A(\omega) \simeq Z_0\,\Re\,\sigma_{\rm 2D}(\omega)
= \frac{\Re\,\sigma_{\rm 2D}(\omega)}{\varepsilon_0 c}.
\end{align}
For graphene, where $\sigma_{\rm 2D}=e^2/(4\hbar)$ in the relevant window, this gives
\begin{align}
A \simeq \frac{e^2}{4\hbar\,\varepsilon_0 c} = \pi\,\alpha_{\rm fs}\approx 2.3\%,
\end{align}
as observed in Nair \textit{et. al} (2008)\cite{Nair2008} and discussed in Dressel \& Gurrnel (2002) \cite{DresselGruner2002}.

Finally, in atomic units $c=1/\alpha_{\rm fs}$. Relativistic (Pauli/Dirac) spin--orbit matrix elements scale as $1/c^2=\alpha_{\rm fs}^2$, so responses governed by SOC inherit an additional overall $\alpha_{\rm fs}^2$ factor through their vertices, on top of the universal $e^2/\hbar$ or $e^2/h$ prefactors \cite{DresselGruner2002}. Throughout, $\boldsymbol{\alpha}(\omega)$ is the electronic polarizability (molecular or per cell), while $\alpha_{\rm fs}$ is the universal constant; rewriting
\begin{align}
\frac{e^2}{\hbar}&=4\pi\varepsilon_0 c\,\alpha_{\rm fs},\\
\frac{e^2}{h}&=2\,\varepsilon_0 c\,\alpha_{\rm fs}
\end{align}
uniformly exposes how $\alpha_{\rm fs}$ sets the overall scale of $\sigma(\omega)$, $\varepsilon_r(\omega)$, and $\sigma_{\rm abs}^{\rm mol}(\omega)$ once the dimensionless correlation kernels are fixed by the material.

\section{Numerical results and validation}
\label{sec:res-val}

The numerical tests in this section are designed to validate, point by point, the workflow outlined in Secs.~\ref{sec:intro}–\ref{sec:fineStructDependency}. 
Figures~\ref{fig:F1}–\ref{fig:F2} probe the \emph{gauge bridge} between length-gauge polarizability and velocity-gauge (Kubo) conductivity: they confirm that, once the equal-time (diamagnetic/contact) term is included alongside the paramagnetic current, the two routes to $\varepsilon(\omega)$ coincide across RF–UV and the optical $f$-sum saturates with the correct prefactor. 
Figures~\ref{fig:F3}–\ref{fig:F4} then exercise the \emph{application layer} of the same framework: the sheet vs.\ ultrathin-film comparison tests the dielectric-matrix/local-field treatment at interfaces, while the RF–microwave skin-depth curves test the propagation through the media end of the framework. 
In all cases, the same microscopic inputs (transition dipoles and kernels), the same conventions for $\varepsilon(\omega)$ and $\sigma(\omega)$, and the same SI prefactors (made explicit via $\alpha_{\rm fs}$) are used without retuning between examples.

\subsection{Gauge checks: length–velocity equivalence and \texorpdfstring{$f$-sum}{f-sum}}

We compare two independent formulations of the dielectric function of the same single-oscillator system:
\begin{align}
\varepsilon(\omega) &= 1 + \frac{N\,\alpha(\omega)}{\varepsilon_0}, 
\label{eq:eps_from_alpha}\\
\varepsilon(\omega) &= 1 - \frac{i}{\varepsilon_0\,\omega}\,\sigma(\omega),
\label{eq:eps_from_sigma}
\end{align}
where the minus sign in \eqref{eq:eps_from_sigma} follows our \(e^{+i\omega t}\) convention.

\begin{figure}[t]
  \centering
  \includegraphics[width=\linewidth]{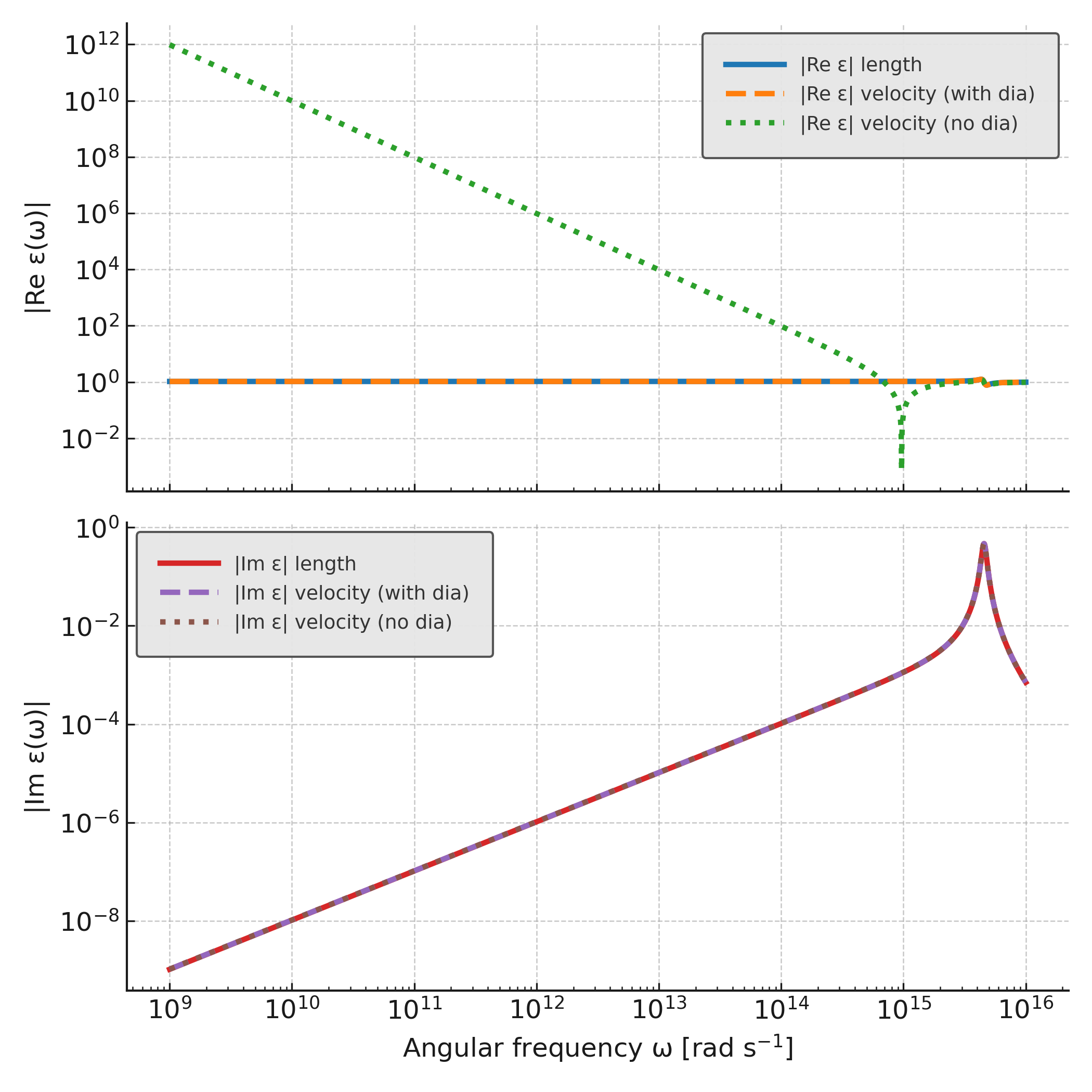}
  \caption{Real (top) and imaginary (bottom) parts of $\varepsilon(\omega)$ for a single Lorentz oscillator, evaluated in two numerically independent ways. The solid curves use the length-gauge polarizability [Eq.~\eqref{eq:eps_from_alpha}], while the dashed curves use the velocity-gauge conductivity [Eq.~\eqref{eq:eps_from_sigma}] including both paramagnetic and equal-time (diamagnetic/contact) contributions. These two routes are numerically indistinguishable across the RF–UV window, demonstrating practical length–velocity equivalence in our implementation. For comparison, the dotted curves show the velocity result when the equal-time term is deliberately omitted (“no-dia”), leading to the expected low-frequency mismatch. All curves are generated from the test-oscillator parameters listed in Table~\ref{tab:allparams}.}
  \label{fig:F1}
\end{figure}

\begin{figure}[t]
  \centering
  \includegraphics[width=\linewidth]{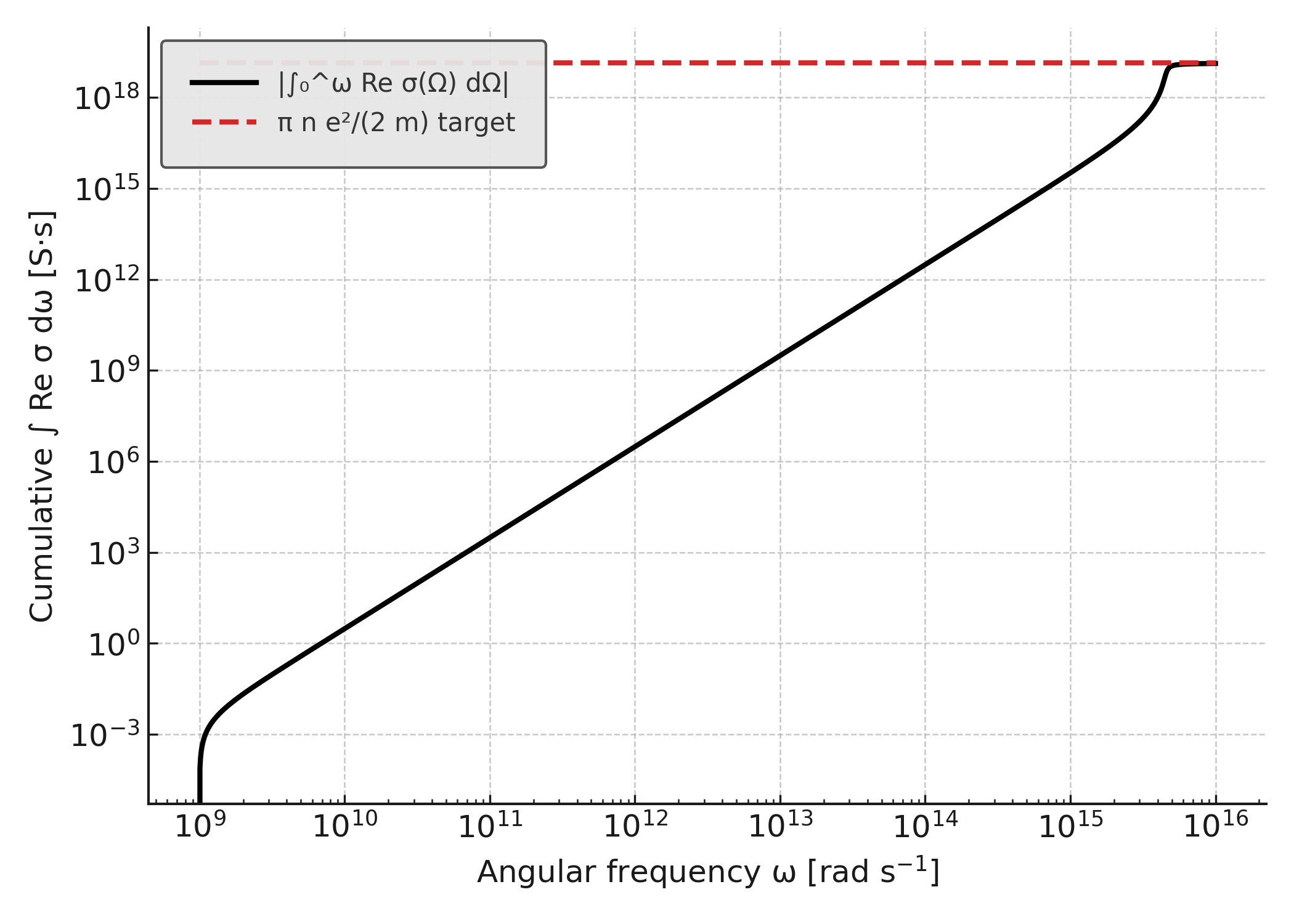}
  \caption{Cumulative integral of the real part of the conductivity,
  $\mathcal{C}(\omega)=\int_0^\omega\mathrm{Re}\,\sigma(\Omega)\,d\Omega$, for the same single-oscillator model as in Fig.~\ref{fig:F1}. The high-frequency plateau coincides with the longitudinal $f$-sum value $\pi n e^2/(2m)$ for this model, to within numerical accuracy. This provides a global check that our velocity-gauge implementation, including the equal-time term and SI prefactors, is internally consistent.}
  \label{fig:F2}
\end{figure}

Figure~\ref{fig:F1} tests the central promise of our framework: that the bridge between dipole-based and current-based response, Eqs.~\eqref{eq:eps_from_alpha}–\eqref{eq:eps_from_sigma}, can be implemented in a way that is \emph{gauge faithful} in practice. 
We consider a single optical oscillator (transition energy $\hbar\Omega$, dipole $\mu$, damping $\gamma$) and compute $\varepsilon(\omega)$ in two completely independent ways.

In the \emph{length-gauge route}, Eq.~\eqref{eq:eps_from_alpha}, we build the polarizability $\alpha(\omega)$ from the usual two-level expression (including dephasing) and then embed a number density $N$ of oscillators to obtain $\varepsilon(\omega)$. This is the standard ``molecular'' construction.

In the \emph{velocity-gauge route}, Eq.~\eqref{eq:eps_from_sigma}, we construct the conductivity $\sigma(\omega)$ using the Kubo formula for the current operator. The current naturally splits into a paramagnetic contribution and an equal-time (diamagnetic/contact) term. The latter is precisely the contribution identified in modern analyses as essential for gauge equivalence but easy to mishandle numerically~\cite{Taghizadeh2017,Taghizadeh2018,Ventura2017,Schuler2021}. 

The two panels of Fig.~\ref{fig:F1} show the absolute values of the real and imaginary parts of $\varepsilon(\omega)$, respectively. In each panel we overlay three curves: the length-gauge result, the velocity-gauge result with both paramagnetic and equal-time terms, and the velocity-gauge result with the equal-time term artificially removed. Across the entire RF–UV span, the length-gauge and full velocity-gauge curves overlay each other within numerical precision, for both real and imaginary parts. This demonstrates \emph{practical} length–velocity equivalence in our implementation: once the equal-time term is consistently included, the two formulations are indistinguishable at the level of $\varepsilon(\omega)$.

In contrast, omitting the equal-time term visibly degrades the result. The ``no-dia'' curves show the familiar unphysical behaviour at low frequency (incorrect static limit and distorted spectral weight), and small but systematic deviations at higher frequencies. The qualitative effect of omitting the equal-time term is consistent with the gauge analyses of Refs.~\cite{Taghizadeh2017,Taghizadeh2018,Ventura2017,Schuler2021}, which emphasize that this contribution is essential for length–velocity equivalence in practical calculations. In our test model, these deviations arise without any additional tuning, purely as a consequence of dropping the equal-time piece. This strongly indicates that the equal-time term is being implemented correctly in our code, and that we are not compensating mistakes with ad hoc prefactors.

As an independent global check, we use the same $\sigma(\omega)$ that enters Eq.~\eqref{eq:eps_from_sigma} to monitor the optical $f$-sum. On a logarithmic grid in $\omega$, we accumulate
\begin{align}
\mathcal{C}(\omega) \;=\; \int_{0}^{\omega}\!\mathrm{Re}\,\sigma(\Omega)\,d\Omega,
\qquad 
\lim_{\omega\to\infty}\mathcal{C}(\omega) \;=\; \frac{\pi n e^2}{2m},
\label{eq:fsum}
\end{align}
with $n$ the effective carrier density. 
This is the usual longitudinal optical $f$-sum rule for the conductivity; see, for example, Refs.~\cite{GiulianiVignale2005,DresselGruner2002} for derivations and discussion in the many-body and solid-state contexts.
Figure~\ref{fig:F2} shows that $\mathcal{C}(\omega)$ cleanly saturates at $\pi n e^2/(2m)$ using the same discretization and prefactors as in Fig.~\ref{fig:F1}. The numerical saturation of $\mathcal{C}(\omega)$ at $\pi n e^2/(2m)$ therefore matches the standard textbook behaviour of the optical $f$-sum rule in metals and insulators.\cite{GiulianiVignale2005,DresselGruner2002} Any sign error, missing factor of two, or mismatch between the $\alpha$-based and $\sigma$-based branches of the implementation would either prevent saturation or shift the plateau. Instead, the combined agreement of Figs.~\ref{fig:F1} and~\ref{fig:F2} confirms that our bridge is both \emph{gauge faithful} (length/velocity equivalence including the equal-time term) and \emph{unit faithful} (sum rules and SI prefactors are correct).

\subsection{Interfaces and propagation: local-field mixing and penetration depth}

\begin{figure}[t]
  \centering
  \includegraphics[width=\linewidth]{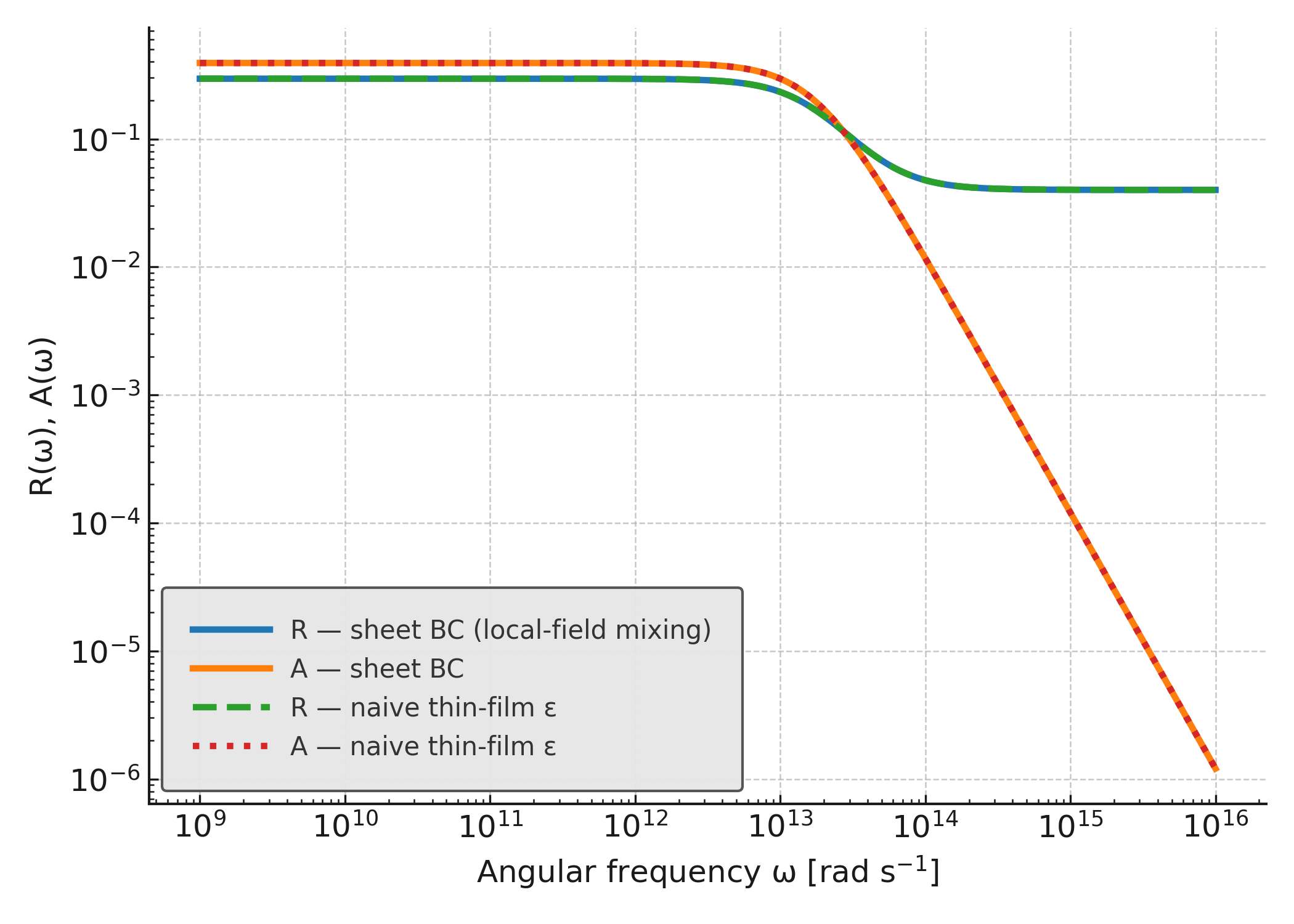}
  \caption{Reflectance $R(\omega)$ and absorbance $A(\omega)$ for a conductive 2D sheet on a dielectric substrate, computed in two equivalent ways. Solid curves: sheet boundary conditions formulated directly in terms of the sheet conductivity $\sigma_{2\mathrm{D}}(\omega)$ in the $q\!\to\!0$ limit. Dashed/dotted curves: ultrathin-film description in which the same sheet conductance is represented by a uniform slab of thickness $d=1$\,nm, with scalar $\varepsilon(\omega)$ obtained from a 3D conductivity $\sigma_{3\mathrm{D}}(\omega)=\sigma_{2\mathrm{D}}(\omega)/d$ and treated as a standard three-layer Fresnel problem. For $d\ll\lambda$ and this matching of conductances, the two descriptions are analytically equivalent to leading order, so the corresponding $R(\omega)$ and $A(\omega)$ curves lie almost on top of each other. Figure~\ref{fig:F3} therefore serves as a numerical consistency test between the sheet and thin-film implementations, using the parameter set in Table~\ref{tab:allparams}.}
  \label{fig:F3}
\end{figure}

\begin{figure}[t]
  \centering
  \includegraphics[width=\linewidth]{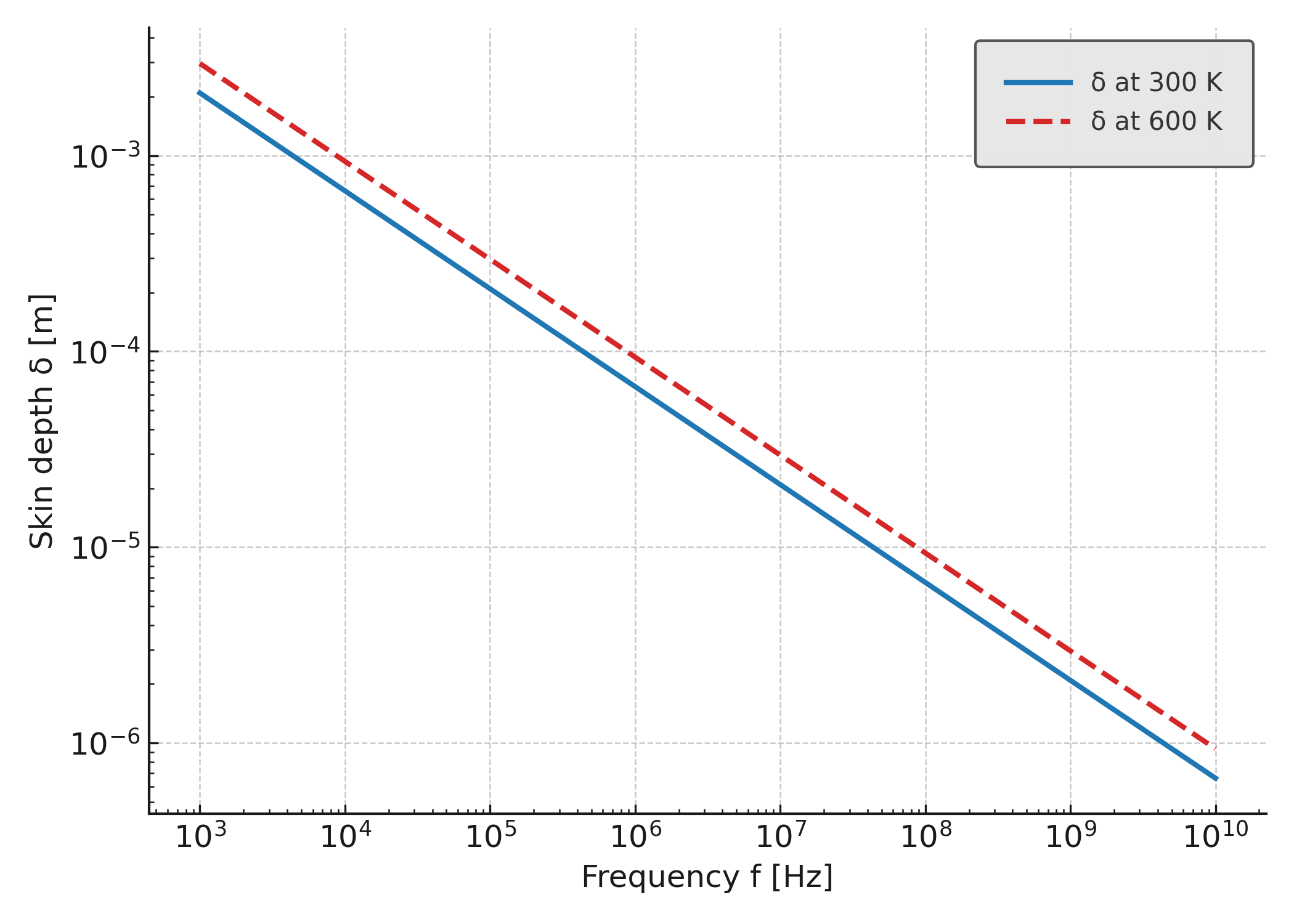}
  \caption{Skin depth $\delta(\omega,T)$ as a function of frequency for two temperatures, computed from a phonon-limited Drude test model. At fixed temperature, $\delta(\omega,T)$ decreases approximately as $\omega^{-1/2}$; at fixed frequency, the sample with larger DC conductivity has a smaller skin depth. These trends are the expected behaviour of electromagnetic waves in good conductors and provide an RF–microwave sanity check for the conventions used in our $\sigma(\omega,T)$ and $\varepsilon(\omega)$ modules.}
  \label{fig:F4}
\end{figure}

Interfaces are where macroscopic optics most visibly feel the microscopic field mixing encoded in the dielectric matrix. Figure~\ref{fig:F3} uses the simplest possible geometry—a single conductive sheet on a dielectric substrate—to verify that our interface formulas reproduce the known ultrathin-film limit. In thin-film optics, it is well known that for $d\ll\lambda$ one can match a 2D sheet conductance to a 3D conductivity $\sigma_{3\mathrm{D}}=\sigma_{2\mathrm{D}}/d$ and obtain the same leading-order Fresnel coefficients as for a sheet boundary condition.\cite{DresselGruner2002} We compare two routes that start from the same sheet conductivity $\sigma_{2\mathrm{D}}(\omega)$.

In the \emph{sheet} formulation (solid curves), the sheet enters Maxwell’s boundary conditions directly through $\sigma_{2\mathrm{D}}(\omega)$, i.e.\ through the $q\!\to\!0$ face of the dielectric matrix. This is the natural language of our framework and keeps the microscopic current response explicit at the interface. In the \emph{ultrathin-film} formulation (dashed/dotted curves), we instead construct a uniform slab of thickness $d=1$\,nm with 3D conductivity $\sigma_{3\mathrm{D}}(\omega)=\sigma_{2\mathrm{D}}(\omega)/d$ and scalar $\varepsilon(\omega)$, and treat this as a standard three-layer Fresnel problem. For $d\ll\lambda$ one can show analytically that these two descriptions are equivalent up to $\mathcal{O}[(\omega d/c)^2]$ corrections.

Consistent with this analytic expectation, Fig.~\ref{fig:F3} shows that the $R(\omega)$ and $A(\omega)$ curves from the sheet and ultrathin-film routes overlay almost perfectly across the plotted frequency window, with only sub-percent differences at the highest frequencies. The near overlap is therefore not an inconsistency but an important sanity check: it confirms that the sheet boundary conditions and the matched thin-film description are implemented with the same conventions and SI prefactors. The situations where the distinction between ``sheet'' and ``naive film'' becomes physically important are more complex systems, such as heterogeneous or anisotropic interfaces.
In these systems, mapping everything to a single scalar $\varepsilon(\omega)$ is no longer justified and local-field mixing between microscopic regions matters; those are the target applications of the general framework developed in Secs.~\ref{sec:peridicSystems} and~\ref{sec:fineStructDependency}.
However, numerical study of such system warrant a careful and in-depth implementation and study which goes beyond the scope of this paper.

Propagation inside conductors is checked in Fig.~\ref{fig:F4}. 
Starting from a simple phonon-limited Drude conductivity $\sigma(\omega,T)$, our macroscopic module computes the RF–microwave skin depth. For a good conductor, the leading-order analysis of Maxwell’s equations gives the standard relation\cite{DresselGruner2002,Mahan2000}
\begin{align}
\delta(\omega,T) \;=\; \sqrt{\frac{2}{\mu_0\,\omega\,\sigma(T)}},
\label{eq:skin_depth}
\end{align}
so that $\delta\!\propto\!\omega^{-1/2}$ at fixed $\sigma$ and $\delta\!\propto\!\sigma^{-1/2}$ at fixed $\omega$. The resulting $\delta(\omega,T)$ curves in Fig.~\ref{fig:F4} follow this behaviour over the frequency range considered and have the correct order of magnitude for metallic conductors. From a workflow perspective, Fig.~\ref{fig:F4} shows that once $\sigma(\omega,T)$ has been fixed at the microscopic level, macroscopic propagation properties follow without any additional ``RF-specific'' conventions or unit surgery.

Taken together, Figs.~\ref{fig:F1}–\ref{fig:F4} show that the proposed long-wavelength TDDFT framework is (i) \emph{gauge invariant in practice} (length–velocity overlay with the equal-time term included), (ii) \emph{unit faithful} (optical $f$-sum and RF skin depth with the correct SI prefactors), and (iii) \emph{interface aware} (sheet boundary conditions and local-field mixing rather than naive ultrathin films). This directly supports the central claim of the paper: that a single, compact bridge from polarizability to Kubo conductivity, built from the same microscopic inputs and $\alpha_{\rm fs}$-explicit prefactors, can be applied unchanged to molecules, bulk crystals, and heterogeneous media from RF to UV.

\subsection*{Reproducibility and units}

All constants and parameters used to generate Figs.~\ref{fig:F1}–\ref{fig:F4} are collected in Table~\ref{tab:allparams}. Throughout we follow the identities \eqref{eq:eps_from_alpha}–\eqref{eq:eps_from_sigma} and the $f$-sum convention in \eqref{eq:fsum}, so that the same set of units and prefactors controls gauge checks (Figs.~\ref{fig:F1}–\ref{fig:F2}), interface optics (Fig.~\ref{fig:F3}), and RF propagation (Fig.~\ref{fig:F4}).
The ``Python'' script to produce the results here presented can be downloaded at \url{https://github.com/Christian48596/tddft-gauge-bridge.git}.

\begin{table*}[t]
  \centering
  \caption{Complete parameter set used to produce Figs. 1--4. Scientific notation $m\times10^{n}$.}
  \label{tab:allparams}
  \renewcommand{\arraystretch}{1.12}
  \begin{tabular}{llclcl}
    \hline\hline
    \multicolumn{6}{l}{\textbf{Physical constants}}\\
    \hline
    Vacuum permittivity & $\varepsilon_0$ & F\,m$^{-1}$ & $8.8541878128\times10^{-12}$ & — & — \\
    Vacuum permeability & $\mu_0$ & H\,m$^{-1}$ & $1.256637062\times10^{-6}$ & — & — \\
    Free-space impedance & $Z_0=\sqrt{\mu_0/\varepsilon_0}$ & $\Omega$ & $3.7673\times10^{2}$ & — & — \\
    Electron charge & $e$ & C & $1.602176634\times10^{-19}$ & — & — \\
    Electron mass & $m$ & kg & $9.1093837015\times10^{-31}$ & — & — \\
    Reduced Planck constant & $\hbar$ & J\,s & $1.054571817\times10^{-34}$ & — & — \\
    \hline
    \multicolumn{6}{l}{\textbf{(A) Two-level oscillator (length \& velocity gauges)}}\\
    \hline
    Transition dipole & $\mu$ & C\,m & $1.000692285\times10^{-29}$ & D & $3.00$ \\
    Transition energy & $\hbar\Omega$ & eV & $3.00$ & Hz & $\Omega/2\pi=7.253967731\times10^{14}$ \\
    Transition frequency & $\Omega$ & s$^{-1}$ & $4.557802346\times10^{15}$ & — & — \\
    Dephasing & $\eta$ & s$^{-1}$ & $2.278901173\times10^{14}$ & $\eta/\Omega$ & $0.05$ \\
    Damping (Lorentz) & $\gamma=2\eta$ & s$^{-1}$ & $4.557802346\times10^{14}$ & $\gamma/2\pi$ [Hz] & $7.253967731\times10^{13}$ \\
    Number density & $N$ & m$^{-3}$ & $1.00\times10^{27}$ & — & — \\
    Lorentz strength & $S=\dfrac{N\,2\mu^2\Omega}{\hbar\varepsilon_0}$ & s$^{-2}$ & $9.776011318\times10^{29}$ & — & — \\
    Effective density (sum rule) & $n_{\mathrm{eff}}$ & m$^{-3}$ & $3.07169884\times10^{26}$ & — & — \\
    Static polarizability (Re) & $\alpha(0)$ & C\,m$^2$V$^{-1}$ & $4.15637830\times10^{-40}$ & cm$^{3}$ (cgs)$^{\dagger}$ & $3.73556653\times10^{-24}$ \\
    \hline
    \multicolumn{6}{l}{\textbf{(B) Interface \& thin-film descriptors}}\\
    \hline
    Sheet DC conductivity & $\sigma_0$ & S & $5.0\times10^{-3}$ & mS & $5.0$ \\
    Scattering time & $\tau$ & s & $1.0\times10^{-13}$ & fs & $1.0\times10^{2}$ \\
    Incident medium index & $n_1$ & — & $1.0$ & — & — \\
    Substrate index & $n_2$ & — & $1.5$ & — & — \\
    Film thickness (naive model) & $d$ & m & $1.0\times10^{-9}$ & nm & $1.0$ \\
    \hline
    \multicolumn{6}{l}{\textbf{Skin depth model}}\\
    \hline
    Reference conductivity & $\sigma(300\,\mathrm{K})$ & S\,m$^{-1}$ & $5.8\times10^{7}$ & — & — \\
    Temperatures & $T$ & K & $300,\,600$ & — & — \\
    Relation & $\delta(\omega,T)$ & — & $\sqrt{2/(\mu_0\omega\sigma(T))}$ & — & — \\
    \hline
    \multicolumn{6}{l}{\textbf{Numerical grids \& integration}}\\
    \hline
    Frequency grid (optical) & $\omega$ & s$^{-1}$ & $\log$-spaced $[10^{9},\,10^{16}]$ & points & $2000$ \\
    Frequency grid (RF) & $f$ & Hz & $\log$-spaced $[10^{3},\,10^{10}]$ & points & $800$ \\
    Conductivity integral & $\int\!\mathrm{Re}\,\sigma\,d\omega$ & — & trapezoidal, $\omega$ grid above & target & $\pi n e^2/(2m)$ \\
    \hline\hline
  \end{tabular}

  \vspace{6pt}
  \footnotesize
  \noindent $^{\dagger}$Conversion used: $\alpha_{\mathrm{SI}} = 4\pi\varepsilon_0\,\alpha_{\mathrm{cgs}}$; $1\,\mathrm{cm}^{3}=10^{-6}\,\mathrm{m}^{3}$; $1\,\mathrm{D}=3.33564095\times10^{-30}$\,C\,m.
\end{table*}

%%%%%%%%%%%%%%%%%%%%%%%%%%%%%%%%%%%%%%%%%%%%%%%%%%%%%%%%%%%%
\section*{Conclusions}
%%%%%%%%%%%%%%%%%%%%%%%%%%%%%%%%%%%%%%%%%%%%%%%%%%%%%%%%%%%%
We presented a compact, gauge–invariant bridge that carries the \emph{same} microscopic inputs—transition dipoles and interaction kernels—across molecules, crystals, and heterogeneous media, with explicit SI prefactors (including $\alpha_{\rm fs}$) and native finite–temperature handling. The construction treats the long–wavelength limit via the dielectric matrix, places the equal–time (diamagnetic/contact) term on the same footing as the paramagnetic current, and keeps units consistent end to end.

The numerical evidence supports both correctness and practicality. In Fig.~\ref{fig:F1}, the dielectric functions from the length and velocity routes become indistinguishable once the equal–time term is included; the divergence that appears when it is omitted is precisely the failure our framework avoids. The optical $f$–sum in Fig.~\ref{fig:F2} saturates in the high frequency limit at $\pi n e^2/(2m)$ to numerical tolerance, auditing our prefactors and confirming that the two gauges use identical microscopic inputs. Interfacial optics in Fig.~\ref{fig:F3} shows why the sheet (dielectric–matrix) treatment is required at $q\!\to\!0$: reflectance/absorbance differ from a naive ultrathin scalar-$\varepsilon$ film, quantifying local–field mixing at boundaries. Finally, Fig.~\ref{fig:F4} links the same inputs to RF–microwave penetration depth with the textbook $\delta\!\propto\!\omega^{-1/2}$ and $\delta\!\propto\!\sigma(T)^{-1/2}$ scalings, demonstrating a unit-faithful pipeline from microscopic kernels to macroscopic observables.

Two practical takeaways generalize across applications. First, many electromagnetic observables can be read as a universal prefactor (set by fundamental constants such as $\alpha_{\rm fs}$) multiplying a nearly dimensionless materials kernel. Framing results this way keeps unit consistency and scaling transparent from bulk 3D solids to strictly 2D sheets; familiar limits, such as the $\pi\alpha_{\rm fs}$ absorption benchmark for graphene, then appear immediately as checks rather than special cases. Second, for realistic surfaces, simple, physics-motivated adjustments to dispersion kernels improve adsorption and wettability energetics while fitting naturally within our workflow: for metals, using the noble-gas $C_6$ above the element with a short real-space cutoff to mimic itinerant screening; for ionic minerals, reducing cation $C_6$ toward noble-gas values to reflect charge transfer to anions \cite{Andersson2013JTC,Andersson2016PCCP,Ataman2016a,Ataman2016b}. These tweaks, calibrated against temperature-programmed desorption, microcalorimetry, QCM-D, and contact-angle data, slot directly into the kernel view we advocate.

Altogether, the bridge turns gauge balance and local-field mixing from abstract consistency conditions into actionable implementation rules. Because it is unit-consistent and gauge-agnostic by design, laboratory spectra and first-principles outputs become directly predictive for kHz–UV applications—heating and penetration depth, dielectric-logging contrasts, thin-film and 2D interfacial optics, and adsorption via imaginary-axis polarizabilities. Immediate priorities include compact temperature-/salinity-aware kernels with quantified uncertainties and extending the interface treatment to small but finite $q$, with clear paths to \emph{operando} diagnostics and digital-twin integration.

\bibliographystyle{aipnum4-1} 
\bibliography{main}

\end{document}